\documentclass{emulateapj}
\usepackage{apjfonts,psfig}

\begin{document}
\title{Gemini/GNIRS Observations of the \\Central Supermassive Black Hole in Centaurus A}
\author{Julia D. Silge and Karl Gebhardt}
\affil{Department of Astronomy, University of Texas at Austin,
    1 University Station C1400, Austin, TX 78712\\{\tt dorothea@astro.as.utexas.edu, gebhardt@astro.as.utexas.edu}}
\author{Marcel Bergmann}
\affil{NOAO Gemini Science Center (Chile), P.O. Box 26732, Tucson, AZ 85726\\{\tt mbergmann@noao.edu}}
\author{Douglas Richstone}
\affil{Department of Astronomy, University of Michigan at Ann Arbor, 
830 Dennison, 501 East University Avenue, Ann Arbor, MI 48109\\{\tt dor@astro.lsa.umich.edu}}

\begin{abstract}

The infrared spectrograph GNIRS on Gemini South unlocks new
possibilities to study the central black holes in dusty
galaxies that have been inaccessible to previous black hole
studies. We exploit good near-infrared seeing to measure the
central black hole (BH) of Centaurus A (NGC 5128). We measure the
stellar kinematics of NGC 5128 using the region around the
CO bandheads at 2.3 $\mu$m and determine the black hole
mass using orbit-based models. 
Black holes are believed to
be essential components of galaxies, and their evolutionary
states appear to be closely linked to those of their hosts.
Our current knowledge does not go much beyond this; galaxies
such as NGC 5128 (an AGN and recent merger) can further develop
this knowledge.
However, NGC 5128 and galaxies like it contain
large amounts of dust which hamper optical spectroscopy,
making near-infrared measurements an attractive alternative.
We find a BH mass of $2.4^{+0.3}_{-0.2}\times 10^8$ M$_\odot$
for an edge-on model,
$1.8^{+0.4}_{-0.4}\times 10^8$ M$_\odot$ 
for a model with inclination of 45$^\circ$, and
$1.5^{+0.3}_{-0.2}\times 10^8$ M$_\odot$ for a model with
inclination of 20$^\circ$.
We adopt the value for the edge-on model, 
which has significantly lower $\chi^2$.
These estimates are consistent with a previous gas dynamical estimate
and are five to ten times higher than that predicted
by the correlation between BH mass and velocity dispersion.
If NGC 5128 will eventually follow the trend for quiescent 
galaxies, this result
suggests that its BH assembled first before its host component.
NGC 5128 thus provides an important example for our
knowledge of central black holes; this technique can
be applied to other such galaxies to further explore
this question.

\end{abstract}

\keywords{galaxies: active --- galaxies: individual (NGC 5128) ---
galaxies: kinematics and dynamics}

\section{Introduction}

NGC 5128 (Centaurus A) is an important object for our
understanding of central black holes, galaxy mergers, 
AGN activity, and the relationships among these components
of galaxy evolution. 
However, the very characterstics which make this galaxy 
so interesting
have also been serious roadblocks to more detailed knowledge.
NGC 5128 contains large amounts of dust which hamper optical
spectroscopy, especially in the central regions which are so
critical for accurately measuring the black hole (BH) mass.
We measure the stellar kinematics of NGC 5128 using new data from the
Gemini Near Infrared Spectrograph (GNIRS) at Gemini South; 
we utilize the region around the CO bandheads at 2.3 $\mu$m.
This observational treatment opens up a new avenue for black 
hole research as it allows us to probe the most interesting 
galaxies; NGC 5128 is the prime example.

NGC 5128 is our nearest neighbor galaxy
harboring a powerful central AGN.  It is a massive elliptical
galaxy which hosts a strong, variable X-ray/radio source
and a massive, complex disk composed of
dust, gas, and young stars.
The unusual morphology of NGC 5128 was first ascribed
to a significant and recent merger event 
by \citet{baa54}.  This merger hypothesis is well-supported
by the existence of optical and HI shells at large radii
and the polar orientation of the disk of dust and gas along
the photometric minor axis of the galaxy \citep{mal83,qui93,isr98}.
The recent merger activity and central AGN are likely
associated with each other, and make NGC 5128 an interesting
case for BH studies. Its notable proximity makes
it an attractive target, as spatial resolution on the sky
translates to small linear scales in the galaxy itself.

The correlation \citep{geb00a,geb00b,fer00} in bulge galaxies 
between central black hole mass (a local property) and 
velocity dispersion (a global property) sheds light on the 
formation and evolutionary histories of both the black hole and its host.
Many theories \citep{sil98,hae00,ost00,kin03} predict such a 
correlation, and with the many models that have been presented 
to date, none have been excluded. One of the best ways to 
determine the underlying physics is to study those galaxies 
that have an active nucleus, i.e. have a central BH which is
actively accreting material, such as NGC 5128. If these galaxies 
lie in a different regime in correlation studies, we can begin
to understand the governing processes and roles of bulge and BH growth.
\citet{geb00b} use reverberation mapping estimates of the BH masses
of AGN galaxies and find the same correlation as for quiescent
galaxies.  NGC 5128 is much closer than any of these galaxies and 
thus we can make a more precise measurement of the BH using
stellar kinematics, the technique used for normal local galaxies.
Also, NGC 5128's recent merger history holds implications for 
the BH-$\sigma$ correlation; we can extend the parameter space
within which we know how galaxies behave in this correlation.

Although galaxies with active nuclei have been a large motivation
in the search for BHs, it is difficult to make a direct dynamical
mass determination for the BHs which we understand to power these
AGNs. The nuclei of many AGN galaxies are heavily obscured by dust,
and NGC 5128 is no exception.  At optical wavelengths, the central
nucleus is nearly invisible, veiled by the rich dust lane.  Such
dust obscuration hampers kinematic measurements made using optical
data; the central BH of a galaxy like NGC 5128 cannot be measured
using optical data. Moving to the near-infrared allows us to
minimize these problems.  Near-IR wavelengths
are long enough to minimize dust extinction.  Also, galaxy light
in this spectral regime is dominated by the older, redder stellar
population and is less affected by recent star formation.
 Kinematics in this spectral
regime should be the best measure of the underlying stellar
potential of the galaxy.  As infrared instrumentation (such as
GNIRS on Gemini South used in this paper) becomes more
available and efficient, this region is becoming an important
tool in the study of galaxies \citep{sil03}.

With such a motivation, \citet{mar01} provide an 
estimate of the BH in NGC 5128 
using $J$-band and $K$-band gas dynamical measurements. 
They find a black hole mass of $2\times10^8$ M$_\odot$. 
Given a velocity dispersion for NGC 5128 of $\sim$150 km s$^{-1}$, 
we would expect a BH mass around $3\times10^7$ M$_\odot$ from
the BH-$\sigma$ correlation. If this BH mass is correct, NGC 5128
has the largest offset ever measured from the BH-$\sigma$ correlation
(currently measured for over forty galaxies).  This is an important
point.  One issue is the difficulty of interpreting gas dynamics.
\citet{sar02} has shown that without 
a well-ordered gas disk it may be impossible to 
determine the enclosed mass from such observations.
\citet{mar01} claim they see no evidence for strong
nongravitational motions, but there are few galaxies with reliable
enclosed masses from both gas and stellar kinematics so it is
difficult to know how to interpret these results.

Thus, NGC 5128 is important because it has recently undergone a merger,
it contains a rich gas disk, its apparently large BH is actively
accreting material, and it is on our doorstep.  In this paper, we
report the BH mass measured from near-infrared stellar kinematics
and its offset relative to other galaxies.
This data can help us refine our knowledge of how galaxies grow 
both their bulge and BH. Section 2 presents the data, Section 3
describes the dynamical modeling procedure, Section 4 presents
the results for the BH mass for NGC 5128, and Section 5 discusses
the implications of these results.

The distance to NGC 5128 is a matter of some debate.  \citet{isr98}
compiles and summarizes results from globular cluster and 
planetary nebulae counts, globular cluster surface brightness
fluctuations, and {\it HST} observations of halo red
giant branch stars; he finds good agreement between these sources
with a distance of $D = 3.4\pm0.15$ Mpc.  More recently, the $I$-band
surface brightness fluctuation study of the galaxy itself
by \citet{ton01} found $D = 4.2\pm0.3$ Mpc. 
\citet{rej04} measures the Mira period-luminosity relation 
and the luminosity of the tip of the red giant branch to find
$D = 3.84\pm0.35$ Mpc. The BH measurement
of \citet{mar01} described above assumes $D = 3.5$ Mpc; we also
use this assumption. At this distance, 
1$^{\prime\prime}$ on the sky corresponds to 17 pc.


\section{Data}

\subsection{Surface Brightness Profile}

To measure the BH mass, we need both photometric measurements
and kinematic measurements with sufficient spatial resolution
and radial extent.  We combine {\it HST} and 2MASS imaging of NGC 5128
to satisfy these needs.  $K$-band {\it HST} observations of NGC 5128
were obtained by \citet{sch98}.  The nuclear region of NGC 5128 was
observed on 11 August 1997 in the F222M filter with an exposure
time of 1280 seconds.  \citet{sch98} report the azimuthally averaged
surface brightness out to a radius of $\sim 10^{\prime\prime}$.
Within $0.65^{\prime\prime}$, the surface brightness profile of
NGC 5128 is dominated by emission from a strong unresolved
source, the central AGN.  We do not include this light
in our dynamical modeling because it does not reflect the stellar
density distribution.  We extrapolate to radii smaller than 
$0.65^{\prime\prime}$ using the {\it HST} data outside this radius.
The logarithm of the surface brightness outside $0.65^{\prime\prime}$ 
is nearly linear with $r^{1/4}$, so we extrapolate this $r^{1/4}$ 
profile inward to our innermost kinematic point.  The dynamical
modeling is not strongly dependent on this extrapolation because
the amount of light (and thus enclosed mass) involved at 
these small radii is not large and does not have a significant impact 
on the gravitational potential.

For photometry at larger radii, we utilize the 2MASS Large Galaxy 
Atlas (LGA) \citep{jar03}.  These authors construct large mosaics for each
of the 100 largest galaxies as seen in the near-infrared.  \citet{jar03}
join 2MASS scans and iteratively remove the sky background,
resulting in carefully constructed, well-calibrated images of 
these galaxies.  We use the
$K$-band image of NGC 5128, the tenth largest galaxy in the atlas.

\psfig{file=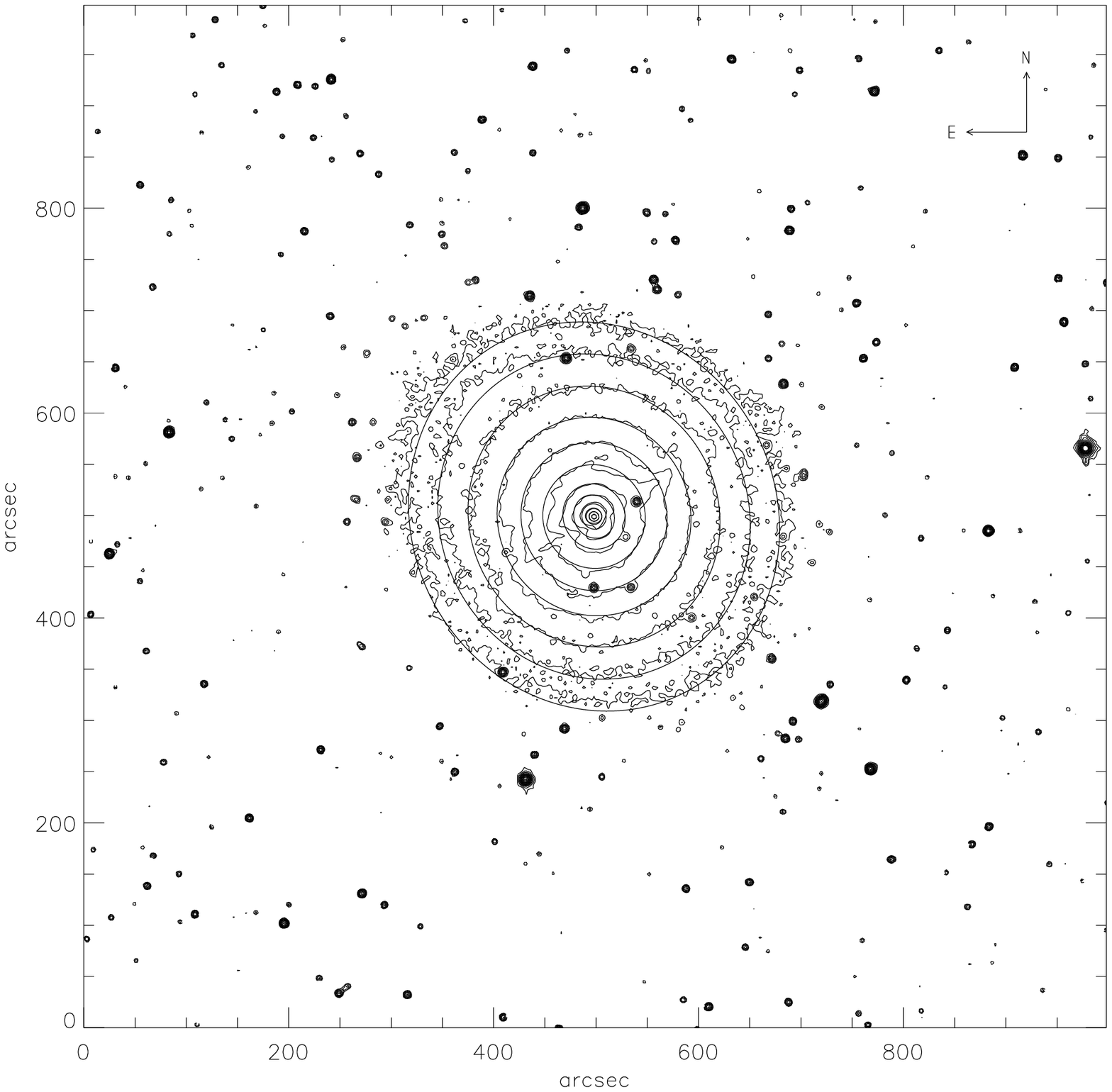,width=8.5cm,angle=0}
\figcaption[CenAcontour_mod.ps]
{Contour map of the 2MASS LGA $K$-band image of NGC 5128.
Overlaid are the contours of the MGE best-fitting model; this model's
profiles are shown in Figure \ref{MGE}.
The contours are logarithmically spaced but arbitrary.
\label{contour}}
\vskip 10pt

We use the multi-Gaussian expansion (MGE) method of \citet{cap02} to fit
the surface brightness profile to the 2MASS image.  The MGE method is
a simple parametrization with an analytic deprojection which is flexible
enough to model realistic multicomponent objects.  This method uses
a series expansion of two-dimensional Gaussian functions to represent 
galaxy images.  Figure \ref{contour} shows the result of MGE fitting for
NGC 5128; this figure shows a contour map of the 2MASS LGA $K$-band image
with the contours of the best-fitting MGE model superimposed.  This model
was constructed to have constant position angle with radius; allowing
the position angle to vary does not improve the fit.
The position angle was fixed at 38 degrees east of north.
The best-fit model uses six two-dimensional Gaussian 
functions to represent the surface brightness of NGC 5128. The
dust lane of NGC 5128 is visible even in this $K$-band image, 
emphasizing the high level of dust obscuration in this galaxy.
The dust lane does not significantly hamper the MGE fitting or the 
kinematic observations below, however.
Figure \ref{MGE} illustrates this; the left panels show the 
comparison between the 2MASS photometry and the best-fitting MGE
model while the right panels show the radial variation of the 
relative error along the profiles.

\psfig{file=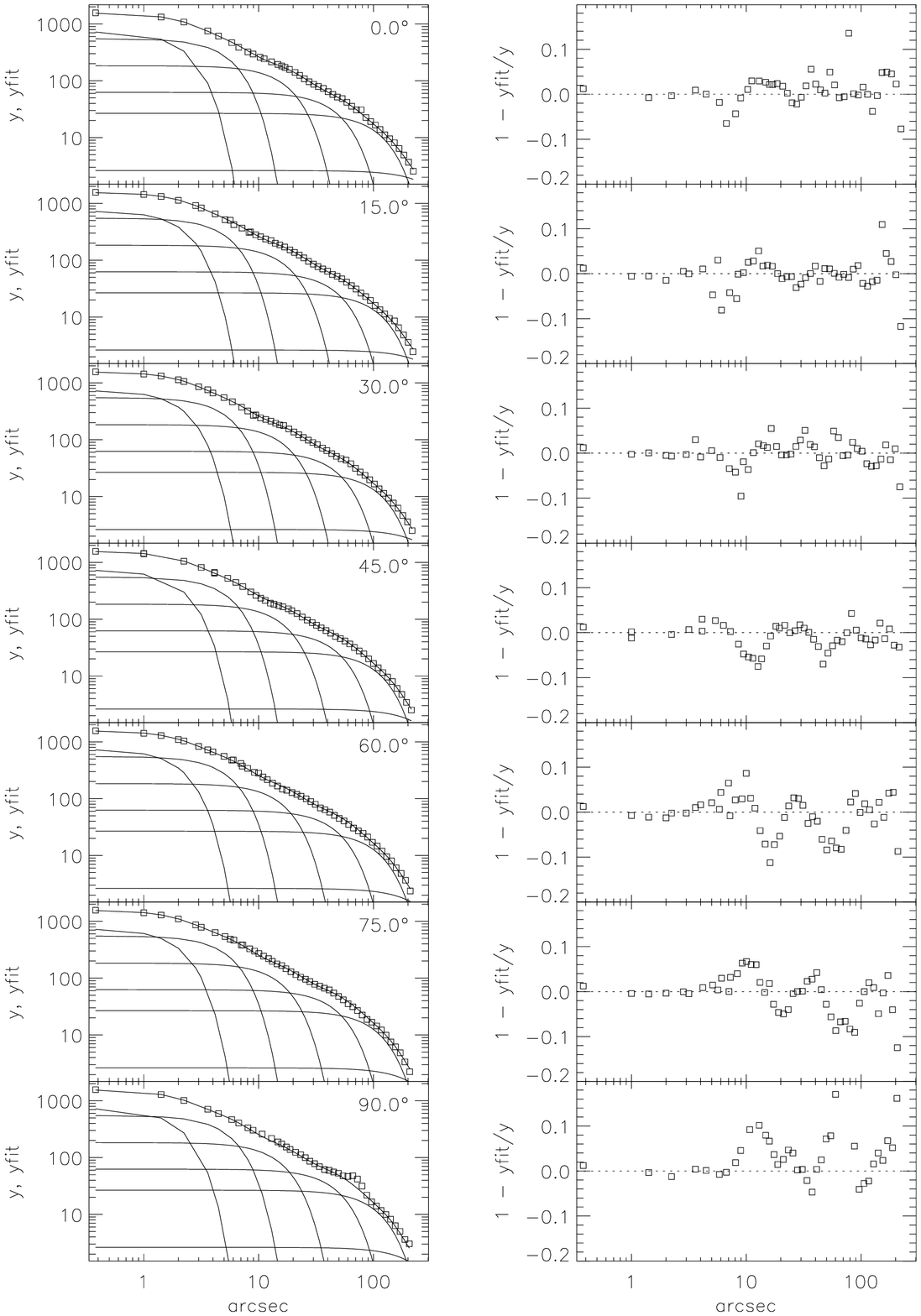,width=8.5cm,angle=0}
\figcaption[N5128fit.ps]{Left panels: comparison between the 2MASS LGA $K$-band
photometry of NGC 5128 (open squares) and the (N=6) Gaussian
MGE best-fitting model (solid line). The individual Gaussian components
are also shown.  The angles noted in the upper right hand corner
of each panel are measured relative to the photometric major axis.
Right panels: radial variation of the relative error
along the profiles.
\label{MGE}}
\vskip 10pt

The dynamical modeling we use to constrain the BH mass assumes
axisymmetry.  Evidence exists that 
NGC 5128 is moderately triaxial \citep{isr98} but the 
photometric data suggest that it can be well-represented for our
purposes as a 
spheroid of constant ellipticity.  The model with constant position
angle fits the surface brightness well, and figure \ref{ellip} shows
that the ellipticity of the galaxy is small and does not change
drastically.  This figure presents the ellipticity of the MGE
best-fitting model as a function of radius; the ellipticity is
never much higher than 0.1.  Thus, we can represent NGC 5128 in our
modeling as a spheroid with constant projected ellipticity of 0.05.  Figure
\ref{SB} presents the final surface brightness profile along
the major axis which we use
in our dynamical modeling.  The dashed line is the profile from 
the MGE best-fitting model of the 2MASS LGA image and the solid line
is the {\it HST} profile of \citet{sch98}.  The {\it HST} data 
have been adjusted
to match the 2MASS data between 2 and 10$^{\prime\prime}$; there
is good agreement in the shape of the two profiles.  The arrow
indicates the radius of transition between domination by the AGN
and domination by the stellar density distribution; the profile
within this radius has been extrapolated from the {\it HST} data outside
this radius.

\psfig{file=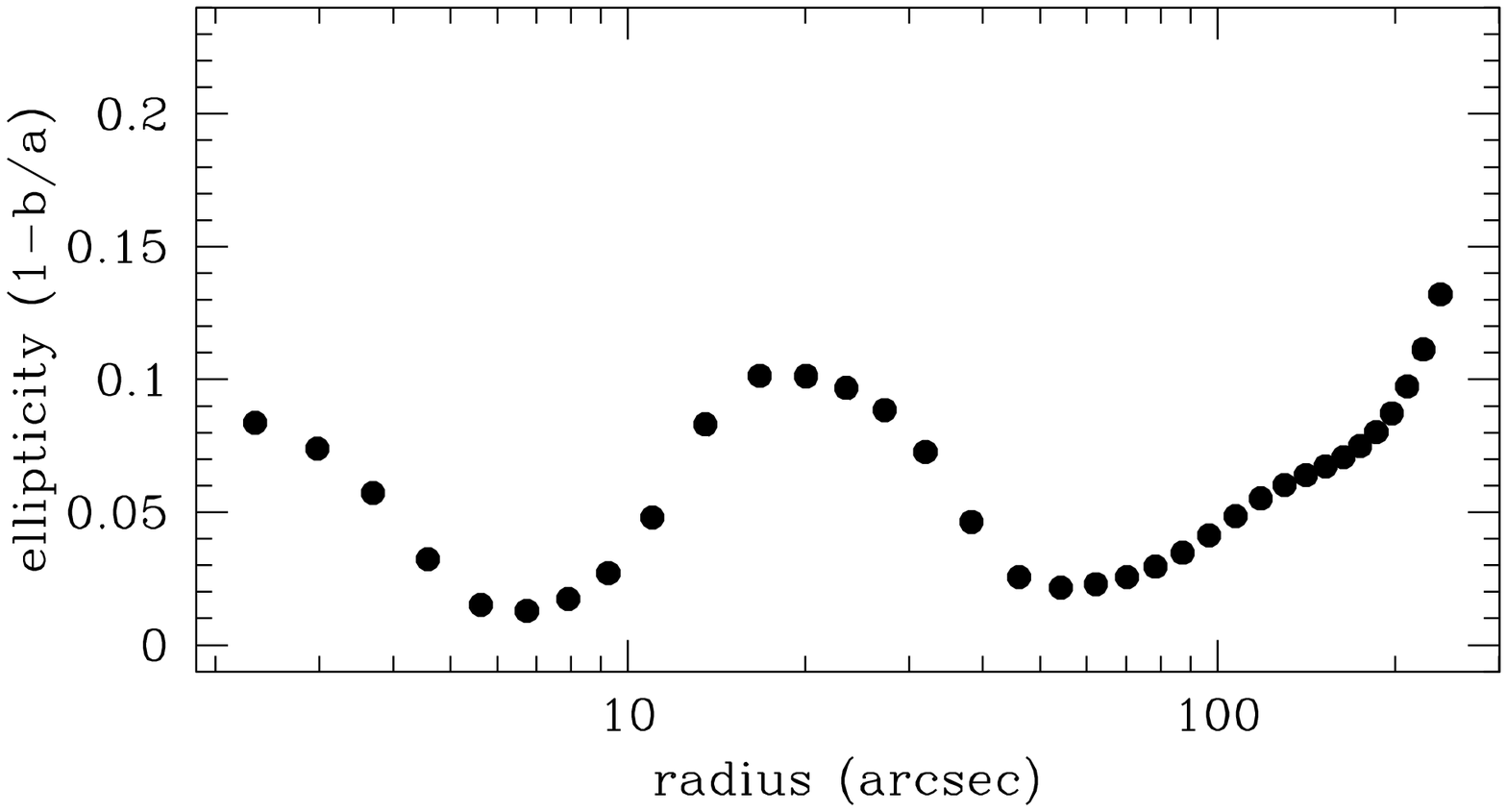,width=8.5cm,angle=0}
\figcaption[ellipticity.ps]
{$K$-band ellipticity as a function of radius along the
major axis for NGC 5128.
\label{ellip}}
\vskip 10pt
\psfig{file=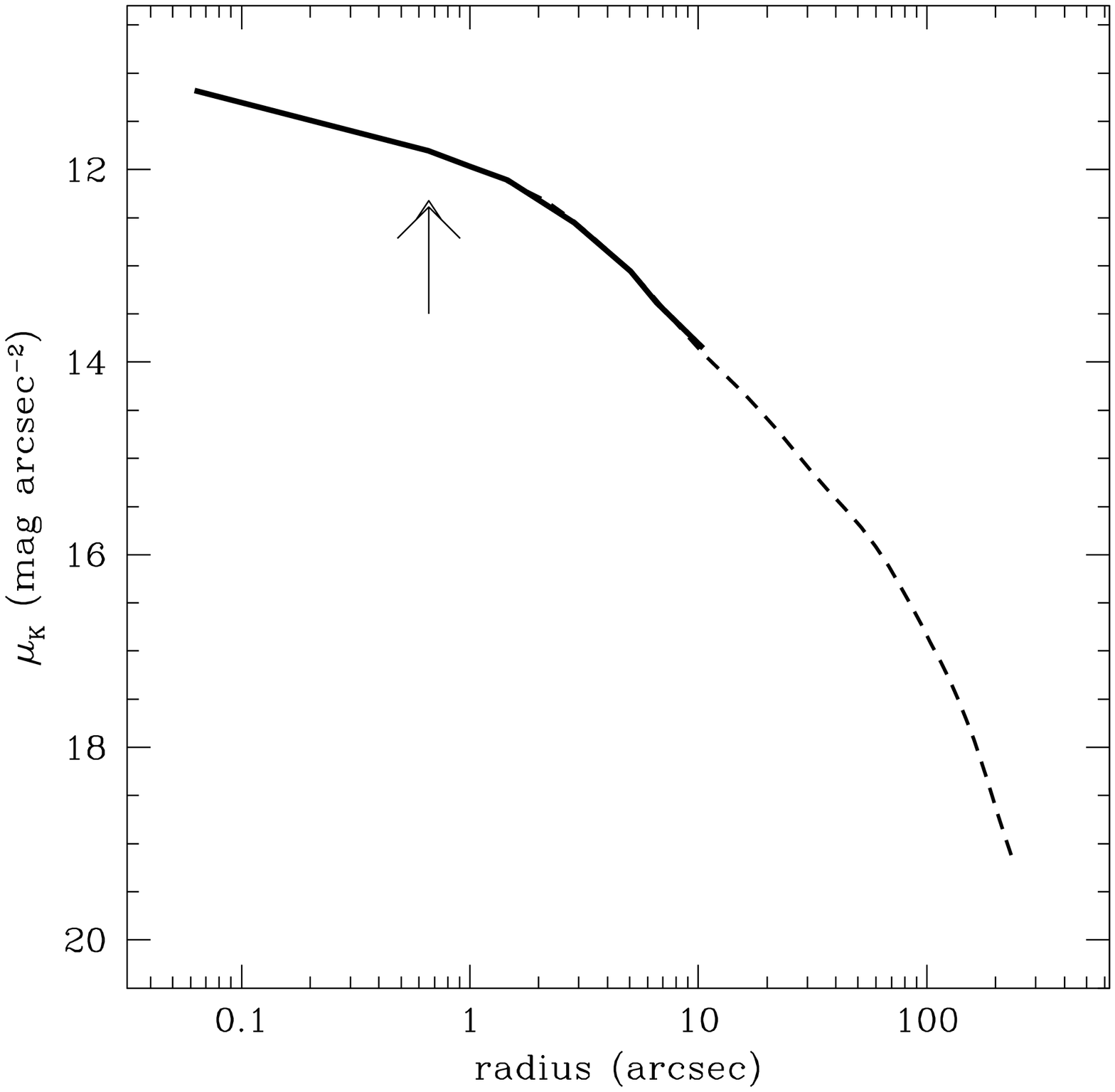,width=8.5cm,angle=0}
\figcaption[CenASBprofile.ps]
{$K$-band surface brightness for NGC 5128.  The dashed line
is from the MGE fitting of the 2MASS LGA image and the solid line
is from the {\it HST} data of \citet{sch98}.  The {\it HST} data 
have been adjusted
to match the 2MASS data between 2 and 10$^{\prime\prime}$.  The arrow
indicates the radius at which light from the AGN dominates; the profile
within this radius has been extrapolated from the {\it HST} data outside
this radius.
\label{SB}}
\vskip 30pt

NGC 5128 is so enshrouded in dust that is reasonable to examine
how affected observations are by dust even in the $K$-band.
\citet{mar00} report {\it HST} observations of the nucleus
of NGC 5128 in $V$, $I$, and $H$, and $K$.  They assume no
color gradients between any of these bands and derive 
the implied reddening for each band.  They report a dereddened
$K$-band surface brightness profile.  The shape of this 
profile is very close
to the observed profile, showing deviation only within 
$\sim1^{\prime\prime}$.  The dereddened profile is still nearly linear
with $r^{1/4}$ so we take the same approach and extend the profile
inward into the region dominated by the AGN light, then compare the mass
implied by the two profiles.  Using the $M/L_K$ ratio implied by the
velocity dispersion profile described in Section 2 (and in agreement with the
dynamical modeling of Sections 3 and 4), the enclosed stellar mass at 
1$^{\prime\prime}$ is $2\times 10^7$ M$_\odot$ for the original profile and
$3\times 10^7$ M$_\odot$ for the dereddened profile.  This difference of 
$1\times 10^7$ M$_\odot$ is a small fraction of the BH mass we measure,
and thus will not have a significant effect.  We ran a small suite of 
dynamical models using the dereddened profile and found that this is indeed
the case; there is no difference in our results.  Similarly, this
applies to the
extrapolation of the stellar light into the region dominated by 
emission from the AGN.  The difference in implied enclosed mass
for different extrapolations is even smaller than for the original
and dereddened profiles and 
the results of our modeling are not dependent
on changes in this extrapolation.

\subsection{Kinematic Observations}

NGC 5128 was observed on 2004 March 8 and June 15 as part of 
system verification for the Gemini Near-Infrared Spectrograph
(GNIRS) on Gemini South, using Gemini program identification number
GS-2004A-SV-8. GNIRS has a 1024 $\times$ 1024 Aladdin III InSb
detector array with a spatial scale of 0$^{\prime\prime}$.15 pixel$^{-1}$.
We utilize the long-slit, short-camera mode of GNIRS which
can obtain the spectrum of the $K$-band atmospheric
window (1.9 $\mu$m to 2.5 $\mu$m) in one exposure.
Using a 32 l/mm grating and 0.3$^{\prime\prime}$ $\times$ 
99$^{\prime\prime}$ slit, we obtained spectral resolution of
1600, measured from calibration lamps lines and night sky lines.
With this resolution, we can measure velocity dispersions down
to 80 km s$^{-1}$, much smaller than the dispersion of NGC 5128.

We utilize the $K$-band CO absorption bandheads 
from evolved red stars to measure the stellar kinematics of NGC
5128. The (2-0)$^{12}$CO absorption bandhead at 2.293 $\mu$m is 
the first of a series of many bandheads which stretch out redward.
These features are in a dark
part of the infrared sky spectrum and are intrinsically sharp and deep,
making this region very sensitive to stellar motions \citep{les94}. 
They are the strongest absorption features in galactic spectra
between 1-3 $\mu$m; this region is
optimal for studying stellar kinematics because
wavelengths are long enough to minimize extinction
from dust but short enough to avoid emission from hot dust \citep{gaf95}.
\citet{sil03} present a detailed calibration of stellar
kinematics measured using the first CO bandhead.

The first observation was made with the slit oriented perpendicular
to the inner dust ring (along the major axis at large radii), 
centered on the bright AGN.  The second
observation was made with the slit oriented parallel to the dust lane
but offset from the center by 0.85$^{\prime\prime}$.  Individual
exposures were 120 seconds; between exposures on the object,
sky exposures of 120 seconds were taken $\sim 200^{\prime\prime}$ 
to the southeast.  NGC 5128 is very large on the sky and large telescope
offsets are required to obtain good sky subtraction.  The total
on-source integration time was 2160 seconds for the slit position 
perpendicular to the dust lane and 1680 seconds for the slit 
position parallel to the dust lane.  We measured the seeing from
images of telluric calibration stars and the central unresolved AGN.
The seeing during the first
observation was 0$^{\prime\prime}$.45; during the second it was
0$^{\prime\prime}$.6.

To remove the shape of the telluric absorption spectrum from
our observations, we observed an A0V star (HD107422).
A dwarfs have nearly featureless spectra in this
region \citep{wal97}; we require a good measure of telluric
absorption because we are interested in the detailed shape
of the galaxy spectrum. \citet{sil03} present more details on this point.
The A0V observations were made by dithering the telescope 
10$^{\prime\prime}$ across the slit to measure the sky at the 
same slit position in alternating exposures. 

Wavelength calibration for both galaxy and star observations
was carried out using the arc lamps of the
Facility Calibration Unit for Gemini South.
Guiding was provided by the peripheral wavefront sensor assigned
to a star outside the science field for the galaxy or telluric
standard.  During the off-source sky exposures for the galaxy
observations, the guiding was paused and we relied on the telescope
tracking because of the difficulty presented by the large offset;
this sky subtraction procedure worked well.
No attempt is made to flux calibrate the spectra since we are only
concerned with the kinematic analysis.

\subsection{Data Reduction}

Data reduction proceeds similarly to that described in \citet{sil03}.
We rectify the images
in the spectral direction using the arc lamp lines and
subtract each sky exposure from its associated object
exposure to remove the sky background.
The subtracted images are then shifted in the spatial direction
so that the center of the galaxy in each image is aligned; we calculate 
the biweight \citep{bee90} of all the processed images to make one 
image for the galaxy.
The one-dimensional spectra are then extracted from the 
two-dimensional image in nearly logarithmically spaced
spatial bins.
The stellar spectra are reduced in a similar manner but extracted
in a single aperture.  
We then remove the telluric absorption spectrum from the galaxy
spectra by dividing by a ``flat'' spectrum, obtained 
from the A0V.  We obtain good results for this spectral
flattening compared to our experience with other instruments and sites.

\psfig{file=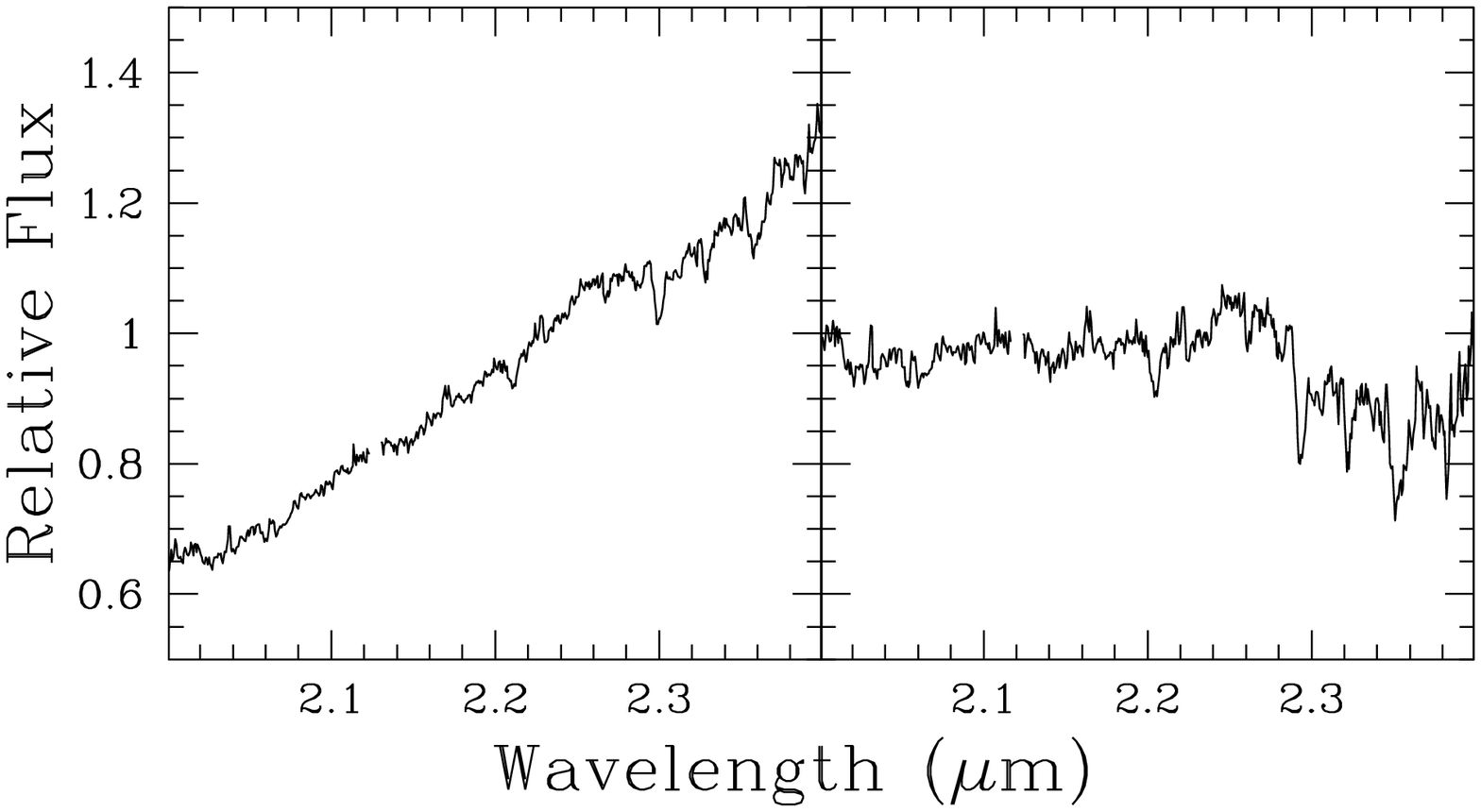,width=8.5cm,angle=0}
\figcaption[AGNremoval.ps]
{Spectrum for one spatial bin 0.$^{\prime\prime}$45 from
the galaxy center.  The left panel shows the spectrum dominated
by the AGN emission; the right panel shows the spectrum after this
AGN continuum shape has been removed.
\label{AGN}}
\vskip 10pt

\begin{figure*}[t]
\centerline{\psfig{file=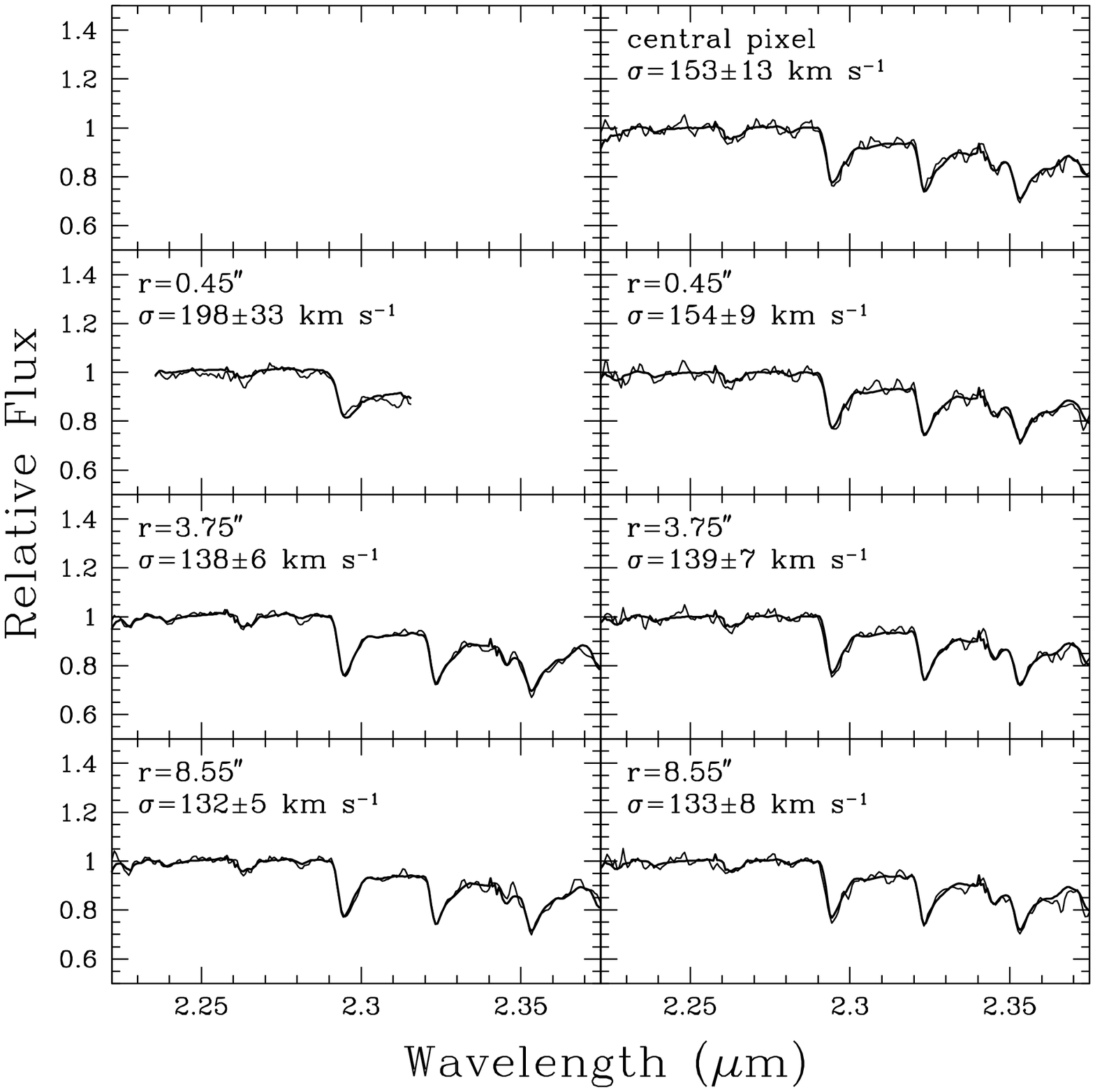,width=14cm,angle=0}}
\figcaption[examplespec.ps]{Rest-frame spectra for seven example
spatial bins (noisy line) and for the template stellar spectrum
convolved with the derived LOSVD for that bin (smooth line).
The derived second moment of the LOSVD and its 68\%
uncertainty is reported for each of these bins. The left
panels show data from the axis perpendicular to the dust disk;
the right panels show data from the axis parallel to the dust disk
but offset from the center. 
\label{examplespec}}
\end{figure*}
The second observation was taken with the slit offset from the center
of the galaxy but the first observation was well-centered on the
central nucleus. The spectra in the central spatial bins 
from this observation are dominated by emission from the central
AGN; the CO bandheads are filled in by this emission.  We worked
to recover some of this information by removing the AGN emission.
The equivalent width of the (2-0)$^{12}$CO
bandhead is largely constant with radius in this galaxy for
regions outside the AGN, so we measure the AGN contribution
by its dilution of the equivalent width.  
After this process, we are able to use data within
0$^{\prime\prime}$.3 of the center in the kinematic fitting.  Spectra
out to 1$^{\prime\prime}$ exhibit changes in equivalenth width
which require removal of AGN light.  Figure
\ref{AGN} illustrates the results of this step for one bin 
0.$^{\prime\prime}$45 from the center.  The left panel
shows the spectrum before removal of the AGN emission and the 
right panel shows the spectrum after this process, ready 
for its kinematic fitting.

\subsection{Extracting the Velocity Distribution}

Once we have the spectrum in each spatial bin, we extract
the kinematic information. A galaxy spectrum is the convolution
of the line-of-sight velocity distribution (LOSVD) 
with an average stellar spectrum; to obtain this internal
kinematic information, we use the fitting technique of 
\citet{geb00}, deconvolving the 
spectrum directly in pixel space using a maximum penalized 
likelihood estimate to obtain a nonparametric LOSVD. An initial velocity
profile is chosen and this profile is convolved with a stellar template
spectrum.  The residuals to the galaxy spectrum are calculated and the
velocity profile is changed to minimize the residuals and provide the
closest match to the observed galaxy spectrum.

The choice of template star is important for kinematic fitting
using the CO bandheads; this feature is sensitive to template
mismatch \citep{sil03}. Thus, we give the fitting program a variety
of template stellar spectra and simultaneously fit for the 
velocity profile and the stellar template weights.
As a result, our fitting procedure provides both the LOSVD 
and stellar population information. We use the near-IR stellar spectral
atlas of \citet{wal97} as our templates, choosing eight stars 
with (2-0)$^{12}$CO equivalent widths ranging from 
less than 5 \AA~to over 20 \AA.  
These spectra have a somewhat higher spectral resolution than
ours, so before using them as stellar templates we have carefully
convolved them to our spectral resolution using a Gaussian
distribution with $\sigma = 5.37$ \AA. 
The best fit
almost always gives most of the weight to a few of the template stars.
Figure \ref{examplespec} shows the results for some example
spatial bins for NGC 5128.  The noisy line is the observed
spectrum and the smooth line is the template stellar spectrum convolved
with the derived LOSVD.  Each frame in figure \ref{examplespec}
shows the spectrum in one spatial bin on one side of the galaxy;
in the actual fitting we fit both sides of the galaxy simultaneously
with the LOSVD flipped around the $v=0$ axis for the opposite side.
For the spatial bins where we removed the AGN emission, we were not
able to use the same wavelength region for the fitting.  The
AGN removal worked best over smaller wavelength regions; in the 
innermost few bins we fit only the first bandhead while in 
intermediate bins we fit the first two.  From 1$^{\prime\prime}$
outward, we used the full region shown in figure \ref{examplespec}.

We determine the uncertainties for the LOSVDs
using the Monte Carlo bootstrap approach of \citet{geb00}.  
The initial fit to the observed galaxy spectrum is used
to generate 100 simulated spectra with noise chosen to match
that of the observed spectrum.
These 100 synthetic galaxy spectra are then deconvolved to
determine their LOSVDs in the same way the original observed
spectrum is deconvolved.
These LOSVDs provide a distribution of
values for each velocity bin which allows us to estimate the uncertainty
and examine any bias in the moments of the LOSVD. 
The median of the distribution 
determines any potential bias from the initial fit, and the spread
of the distribution determines the uncertainty. 
In order to generate the 68\%
confidence bands, we choose the 16\% to 84\% values from the 100
realizations.

Most LOSVD fitting techniques make some assumption about the shape
of the LOSVD, i.e. it is Gaussian or a Gauss-Hermite polynomial.
Our technique obtains a nonparametric LOSVD; no a priori
assumptions about the shape of the LOSVD are made except that it
is nonnegative in all bins and subject to some smoothness constraint. 
We are able to exploit the full
LOSVDs in the dynamical modeling below.  We plot the
parameterization of the LOSVDs by Gauss-Hermite moments in
figure \ref{kinematics} and present them
in table \ref{momenttable}.  The velocity dispersion is very
close to 135 km s$^{-1}$ on both axes for most of
the radius range ($\sim 2^{\prime\prime}$ to $\sim 40^{\prime\prime}$).
The luminosity-weighted $\sigma_*$ using an aperture of 
60$^{\prime\prime}$ along the slit parallel to the dust disk
is 138$\pm$10 km s$^{-1}$.
The axis perpendicular to the
dust lane (which was centered on the galaxy nucleus) shows
a steep rise in both $\sigma$ and $h_4$ (which indicates a
triangular distortion from a Gaussian shape, i.e. strong high
velocity wings on the LOSVD) within the central arcsecond.
We detect rotation along both axes, the photometric major axis
and the axis parallel to the dust disk in the center, as have other
authors including \citet{pen04} and \citet{hui95}.  This is not
strictly consistent with an axisymmetric model, as we discuss later.

These Gemini observations go out to $\sim 40^{\prime\prime}$, but
this is only about halfway to the half-light radius.  According 
to \citet{jar03}, the $K$-band half-light radius is 
$r_{eff}$ = $82.6^{\prime\prime}$.  Limited spatial extent
can significantly reduce the precision to which one can measure
the BH mass \citep{ric04} so we include kinematic data at larger
radii to increase the precision of our measurement.  We use a
kinematic study of planetary nebulae in NGC 5128 \citep{pen04}
which extends well into the galaxy halo.  \citet{pen04} report
the rotation curve and velocity dispersion along the photometric 
major axis out to $80^\prime$, including planetary nebulae within
a perpendicular distance of $\pm120^{\prime\prime}$ to this axis.
The largest radii are not important to us as we are mainly
interested in the gravitational potential at smaller radii; we
include their first three points only, extending
our kinematic coverage to $2r_{eff}$.  These kinematic data
are rotation velocities and velocity dispersions, not full LOSVDs.
In our dynamical modeling, we assume Gaussian LOSVDs with first
and second moments to match the planetary nebulae data for these points.
The lack of higher order information at these large radii does not
effect the BH mass measurement.  The planetary
nebulae data are shown in figure \ref{kinematics} and presented in table
\ref{momenttable}.

\psfig{file=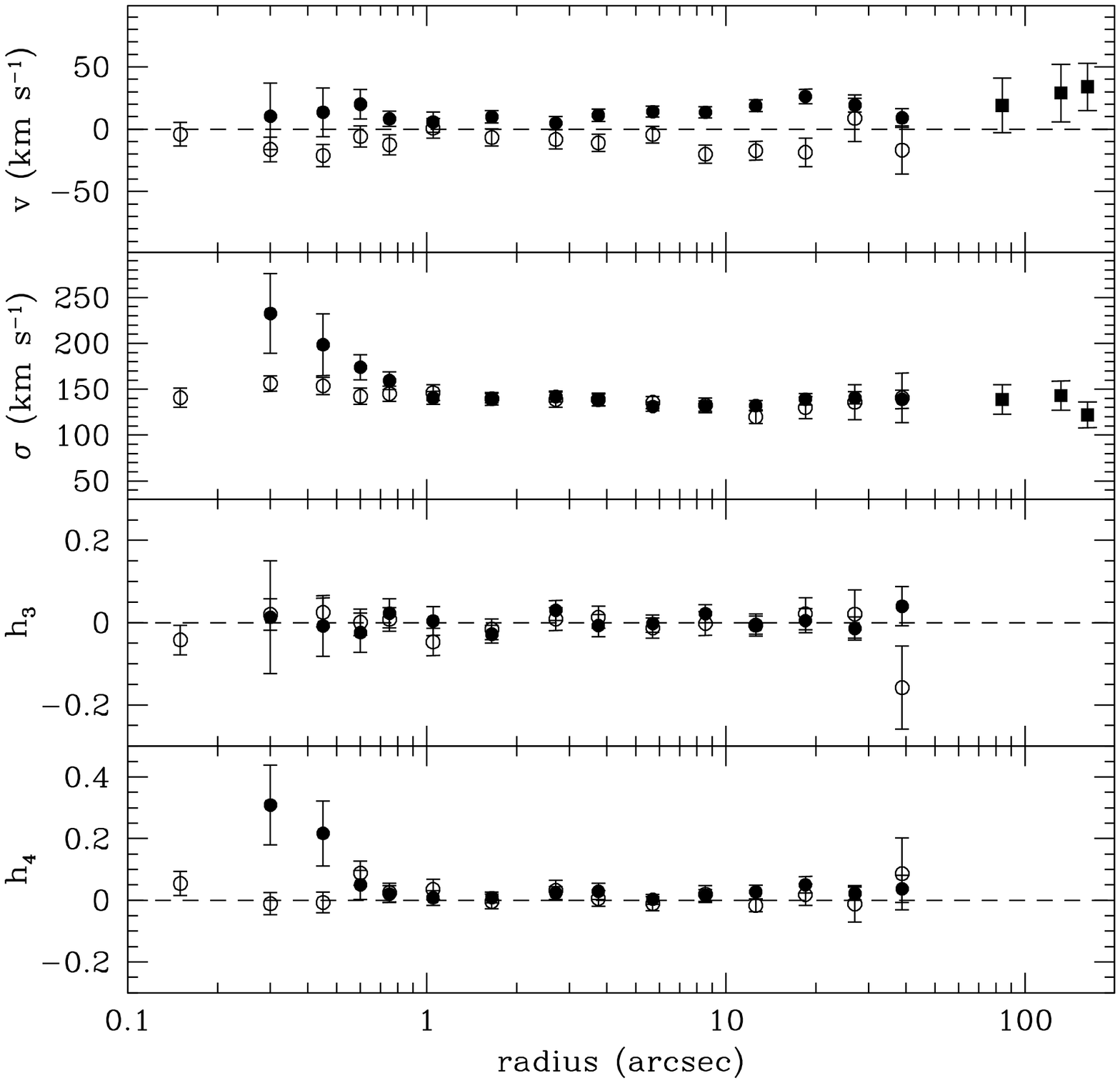,width=8.5cm,angle=0}
\figcaption[CenAmoredata.ps]
{Gauss-Hermite moments of the LOSVDs along the slit 
perpendicular to the dust disk (filled circles) and along the slit
parallel to the dust disk but offset from the center (open circles).
The filled squares are from \citet{pen04} and are along the same axis
as the filled circles.
\label{kinematics}}
\vskip 10pt

\section{Dynamical Models}

The dynamical models are constructed as in \citet{geb00} and
\citet{geb03} using the orbit superposition technique first
proposed by \citet{sch79}.  We use the surface brightness
profile described above to estimate the luminosity density
distribution; the surface brightness can be deprojected
assuming axisymmetry and some chosen inclination. This luminosity
density then translates to a mass density distribution, 
assuming some
stellar mass-to-light (M/L) ratio and BH mass.  It is this mass
density distribution which then defines the potential 
for a given model. Using this derived potential, we follow
a representative set of orbits which sample the available
phase space.  We then determine the orbit superposition
(i.e. nonnegative set of weights for the orbits) which provides
the best match (the minimum $\chi^2$) to the data.
We can impose smoothness on the phase-space distribution function
by maximizing entropy \citep{ric88,tho04}.  We repeat
this process for a variety of BH masses, $M/L$
ratios, and inclinations to find the overall best match.

\begin{figure*}[t]
\centerline{\psfig{file=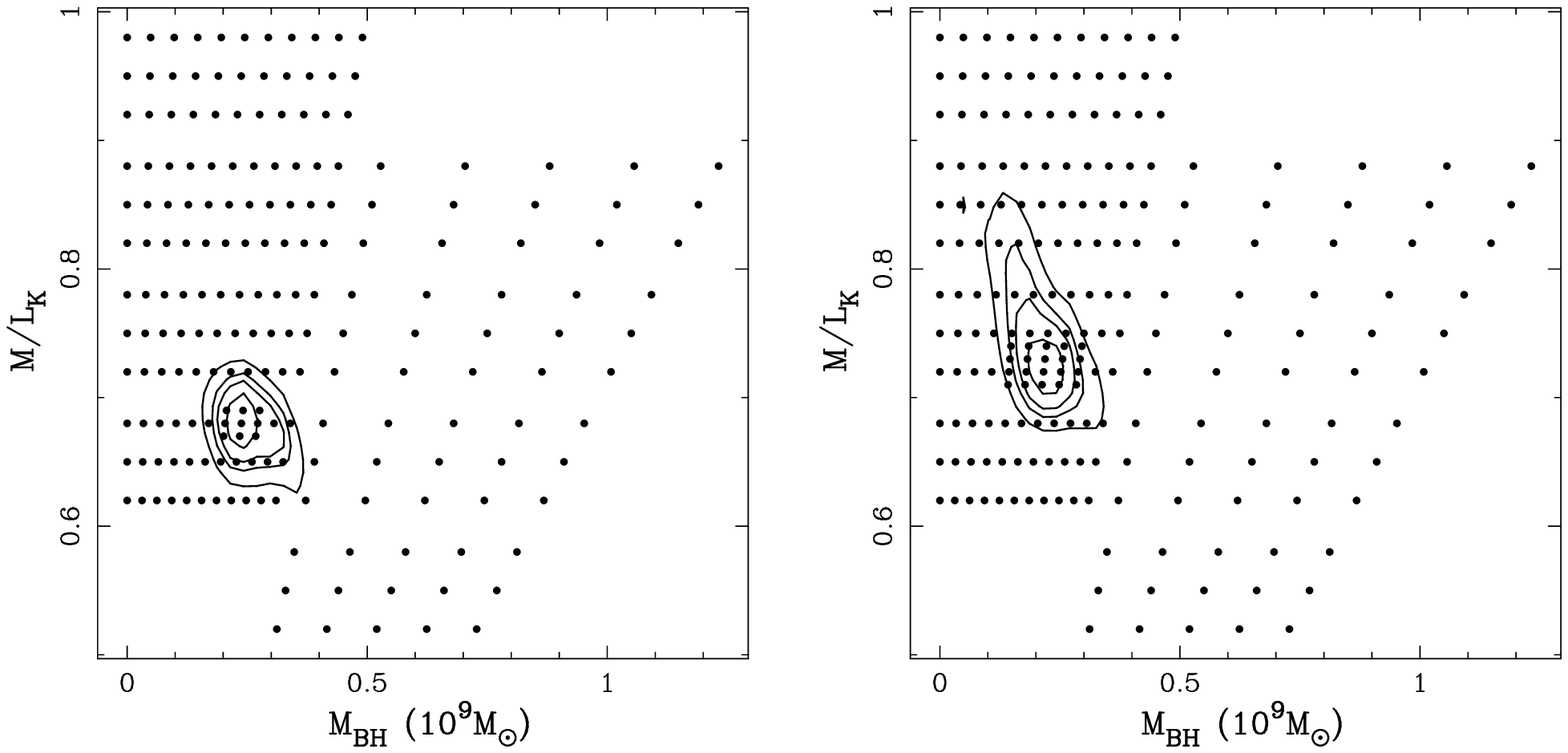,width=16cm,angle=0}}
\figcaption[2dchi.ps]{Two-dimensional plots of $\chi^2$ as a function of 
BH mass and $M/L$ ratio for both orientations we
modeled.  
The left-hand panel shows models with 
rotatation along the galaxy's photometric major axis (PHOT models) 
and the right-hand panel shows models with rotation along
the dust disk (DUST models)
The points represent models. The contours were determined 
by a two-dimensional smoothing spline interpolated 
from these models and represent $\Delta\chi^2$
of 1.0, 2.71, 4.0, and 6.63 
(corresponding to 68\%, 90\%, 95\%, and 99\% for 1 degree of freedom).
\label{BH}}
\end{figure*}

To obtain a smooth $\chi^2$ distribution, it is necessary
to include an adequate number of orbits.  We sample
the gravitational potential by launching orbits from points
in the three-dimensional space spanned by the energy $E$,
the angular momentum $L_z$, and the third integral $I_3$.
The limits of this space are well-understood, and \citet{ric04}
details the considerations for the
sampling of this phase space.  Our final model contains 
$\sim$10,000 orbits,
adequate to obtain a smooth $\chi^2$ distribution for our 
binning scheme.

To compare the model with the data, we map both the observations
and orbits to a grid of 20 radial bins, 5 angular bins, and
17 velocity bins.  The time a given orbit spends in a given
bin translates to its contribution to the light in that bin, and
thus the mass.
The radial binning scheme for the model is
the same as that for the data shown in figure \ref{kinematics}
and table \ref{momenttable};
we have data on the major and minor axes, i.e. in two angular bins
of the model.  The binning schemes are chosen to maximize the
S/N of the data.  The velocity bins span the minimum and maximum
velocities of the orbits; these must be chosen carefully because the
high-velocity wings of the LOSVD are affected the most by a BH.
We incorporate seeing in the model by convolving the light
distribution for each orbit with the PSF of the kinematic
observations.  The Gemini observing runs had PSFs, 
approximated as Gaussians, of 0$^{\prime\prime}$.45 in the March
run and 0$^{\prime\prime}$.60 in the June run.  The planetary
nebulae data have such large binning that 
seeing does not affect them.

For each model, we use the best-fit orbit superposition to match
the light in each of the 100 spatial bins and the LOSVDs in the
33 bins where we have data.  We fit to the full
LOSVDs, not a parameterization such as the Gauss-Hermite moments.
The orbit weights are chosen so that the luminosity density 
in each spatial bin matches the data to better than 1\%, with typical
matches better than 0.1\%. This match is treated as a constraint, not
as a difference to be minimized.  We do minimize $\chi^2$ for each model,
with $\chi^2 = \Sigma [(y_i-y^\prime_i)/\sigma_i]^2$, where the $y_i$'s
are the LOSVD bin heights of the data, the $y^\prime_i$'s are the LOSVD
bin heights of the model, and $\sigma_i$ is the uncertainty of the bin
height of the data.  Each combination of BH mass and stellar $M/L$ ratio
has an orbit superposition which gives a minimum $\chi^2$; we then
compare the $\chi^2$ of different such combinations to find the
best match to the data. 
We use this measure of $\chi^2$ to determine the uncertainties in 
the BH mass and stellar $M/L$ as well.  These are
correlated, as the model can exchange mass in the BH for higher $M/L$,
and thus we use the two-dimensional
$\chi^2$ distribution to determine the 68\% confidence bands
for these quantities, where $\Delta\chi^2 = 1$ for one degree
of freedom.

\section{Results}

The fact that NGC 5128 has rotation along more than one axis
has interesting implications for its dynamical structure.
This galaxy is either tumbling or triaxial.  Tumbling is not
an equilibrium dynamical state, but as a recent major merger,
NGC 5128 may well not be in dynamical equilibrium.
Strictly speaking, this means that we cannot match well
the kinematics of NGC 5128 and the kinematics of an
axisymmetric orbit superposition model.  In our models,
we choose to identify one axis of the observations with
the rotation axis of the model; the rotation on the other axis
serves only to balloon up the $\chi^2$.  We find that the best-fitting
models for NGC 5128 constructed in this way are not very good fits
and have unrealistic characteristics; we choose to take
another approach.

We return to our LOSVDs in each bin and symmetrize them about $v=0$, 
taking the mean of each side to make the new, symmetric LOSVD.
These new LOSVDs have no net rotation.
This is not the true state of NGC 5128, but the resulting
LOSVDs do imply approximately
the same enclosed mass as our ``true'' LOSVDs.
The kinetic energy is proportional to $(v^2+\sigma^2)$; 
using a symmetric LOSVD forces the kinetic energy into the $\sigma^2$ term.
We can then build orbit superposition models to match 
the ``true'' kinematics along one axis which we identify
as the axis of rotation and these no-rotation kinematics
along the other axis.

We are then faced with the quandary of which axis to identify
as the rotation axis.  It seems logical that
the rotation along the direction of the
dust disk at small radii would be important for the determination
of the BH mass; however, inclusion of rotation at large radii also has 
important affects on inferred BH masses \citep{ric04}.
It is unclear what the best choice would be, especially since
BH masses have not been determined using orbit superposition models
for galaxies with
such complex dynamical structure as NGC 5128.  We thus choose to
repeat our entire modeling procedure for both cases, first 
matching to the rotating kinematics along the dust disk axis and non-rotating
kinematics along the photometric major axis (hereafter DUST), 
then switching 
to rotating kinematics along the photometric major axis
and non-rotating kinematics along the dust disk axis (hereafter PHOT).  
We can thus
compare the BH masses inferred by including different data.

Figure \ref{BH} presents the results of this process. Each panel
shows $\chi^2$ as a function of BH mass and $M/L$ ratio,
obtained by comparing
model kinematics to observed kinematic data for NGC 5128. 
The left panel shows the DUST models
and the right panel shows the PHOT models.
The contours are drawn using a two-dimensional smoothing spline
\citep{wah90} but the modeled values are fairly smooth and
large smoothing is not necessary.
Figure \ref{chi2} presents the $\chi^2$ goodness of fit as a
function of black hole mass; we have marginalized 
the two-dimensional $\chi^2$
distributions shown in figure \ref{BH} over $M/L$ to obtain
these distributions.  The dashed line and open squares
are for the DUST models while the
solid line and filled circles are for the PHOT models.
The points show
the $\chi^2$ of individual models; the lines are marginalized from
the smoothed distributions illustrated in the contours shown
in figure \ref{BH}.  The models with rotation along the dust axis
had overall higher $\chi^2$ values so these are offset vertically
for plotting purposes.

\psfig{file=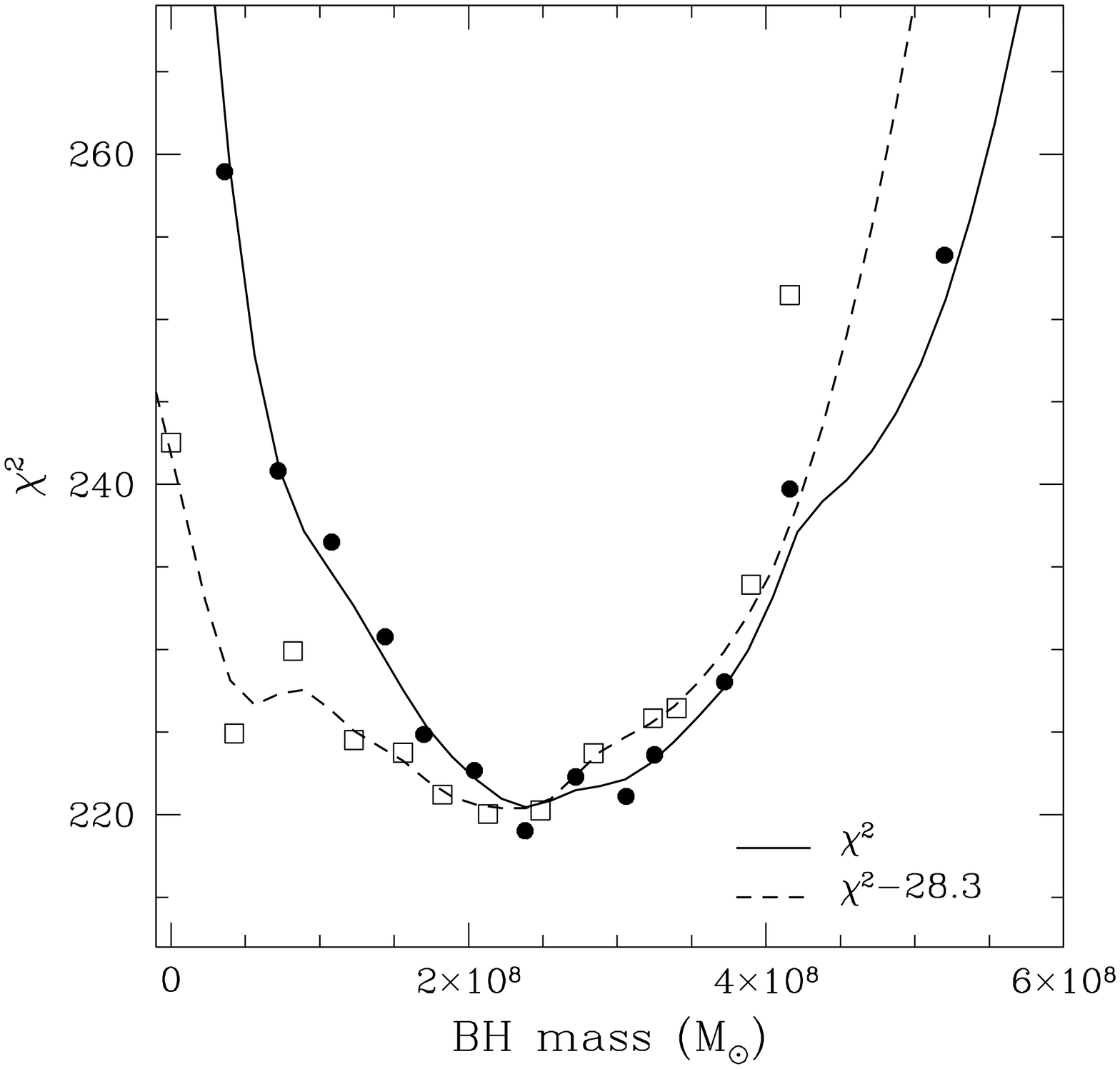,width=8.5cm,angle=0}
\figcaption[BHplot.ps]{The $\chi^2$ goodness of fit obtained by comparing
model kinematics to observed kinematic data for NGC 5128,
versus model black hole mass.  The dashed line and open squares
are for the models with the rotation axis along the dust disk (DUST models),
and the solid line and filled circles are for the models with
the rotation axis along the photometric major axis of the galaxy (PHOT models).
\label{chi2}}
\vskip 10pt

The BH masses from the two sets of models are in excellent
agreement.
For the PHOT models, the BH mass is 
$2.4^{+0.3}_{-0.2}\times 10^8$ M$_\odot$; for the DUST models, it is
$2.2^{+0.3}_{-0.3}\times 10^8$ M$_\odot$.
The PHOT models have significantly lower $\chi^2$ ($\Delta\chi^2=28.3$),
so we identify this as the preferred orientation. The photometric major
axis of the galaxy (the rotation axis of the PHOT models)
is the important axis for most of the galaxy's mass; it is reasonable
that this is more important for a dynamical model of the galaxy.
The $M/L_K$ ratios are $0.68^{+0.01}_{-0.02}$ for the PHOT models 
and
$0.72^{+0.03}_{-0.02}$ for the DUST models.
These are not in agreement to within the
68\% confidence limits but are close.
Interestingly, there is some covariance in the DUST models
but very little in the PHOT models.  The PHOT models use the
data that were well-centered on the galaxy nucleus on the model's
rotation axis.  In contrast, the DUST models use the data that
were offset from the nucleus along the direction of the model's
rotation axis and thus do not have any data on the actual rotation axis
of the model.  Such data appear to be important in breaking the
degeneracy between BH mass and $M/L$ ratio and placing the strongest
constraints on each quantity.

\begin{figure*}[t]
\centerline{\psfig{file=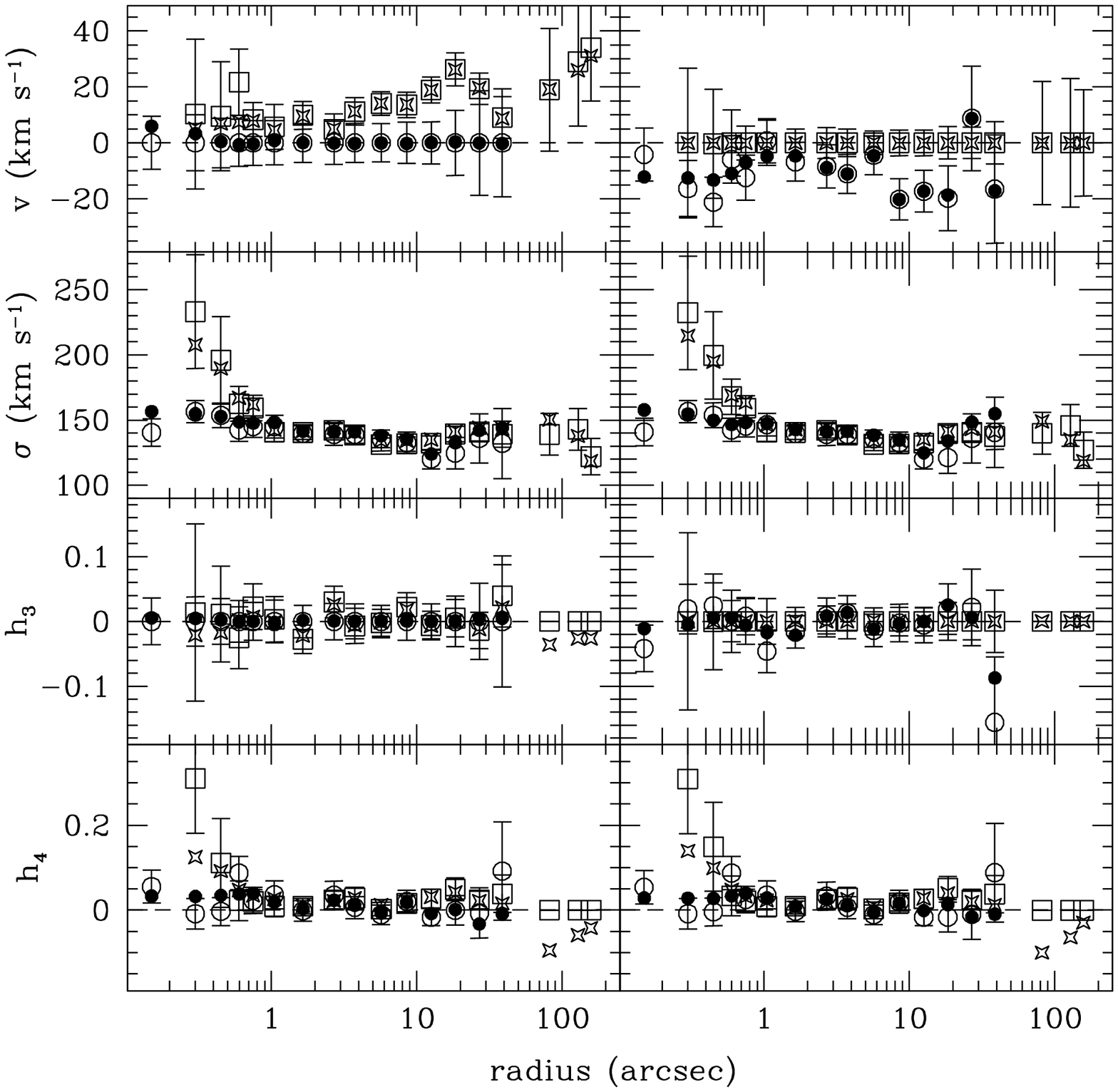,width=12cm,angle=0}}
\figcaption[comparemodel.ps]
{Comparison of observed kinematic data along each slit for NGC 5128
with model kinematics for the two best-fitting edge-on models.
The left-hand panel compares to the best-fitting edge-on model with
rotation along the photometric major axis (PHOT models) and the right-hand
panel compares to the best-fitting edge-on model with rotation along
the dust disk (DUST models). The open symbols are the data: circles for
the axis parallel to the dust disk but offset from the center
and squares for the galaxy photometric major axis.
The filled circles are from the model on the 
axis parallel to the dust disk and the starred squares are
from the model on the galaxy photometric major axis.
\label{compare}}
\end{figure*}

Figure \ref{compare} compares the model kinematics with the
observed kinematics for each axis on which we have data, for
each orientation of rotation that we modeled.  The left panels are
for the PHOT models and the right panels are for the DUST models.
The squares are from the photometric major axis and the circles
are from the axis parallel to the dust disk but offset from the
nucleus.  The open symbols are the actual (or symmetrized) data
and the filled or starred symbols are from the models.
We emphasize that we fit the full LOSVDs and that this plot
shows only the Gauss-Hermite moments of the LOSVDs.
Notice the symmetrization of the observed data;
$v=0$ for every radial point on the dust axis for the PHOT
models and every radial point on the photometric major axis
for the DUST models.

\begin{figure*}[t]
\centerline{\psfig{file=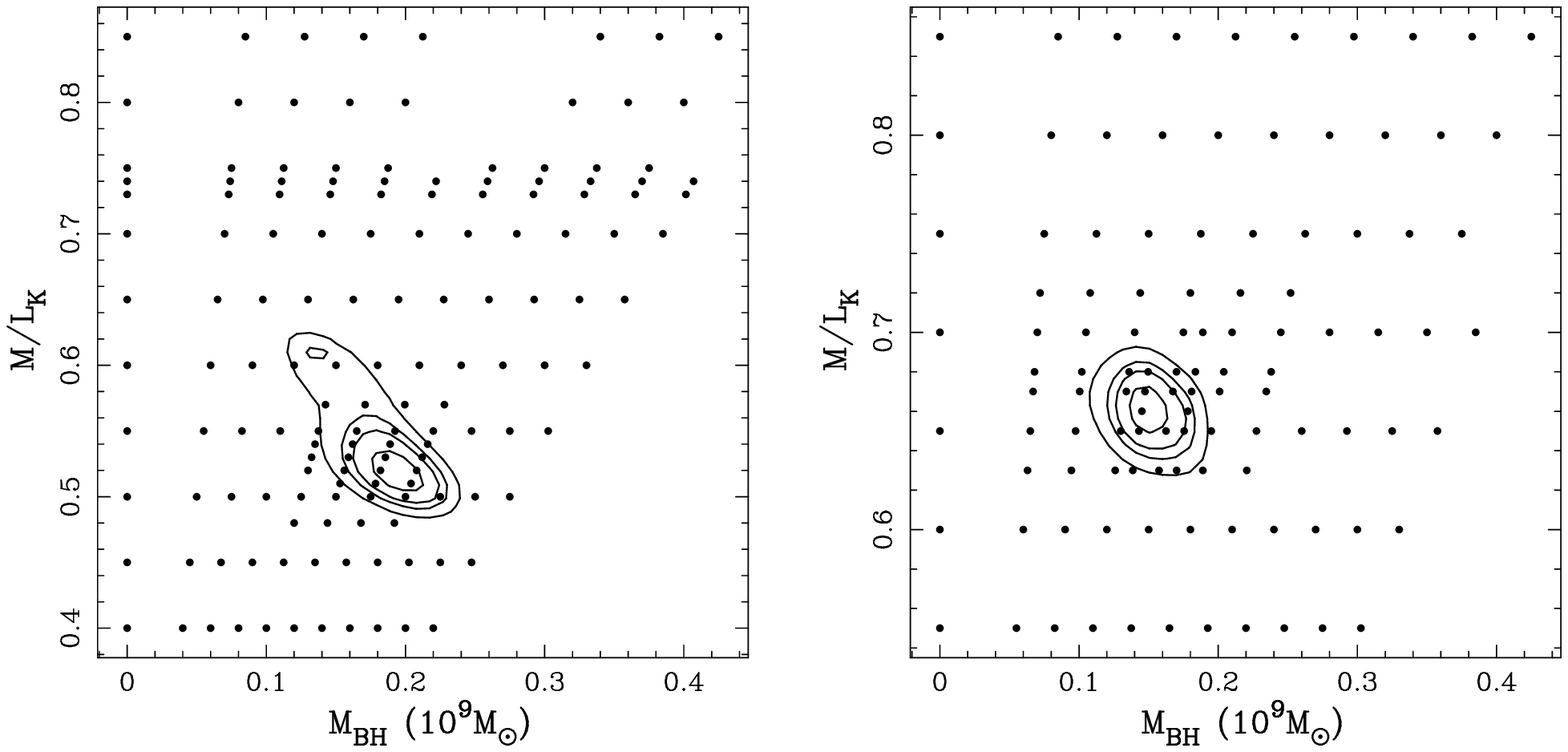,width=16cm,angle=0}}
\figcaption[2dchiincline.ps]{Same as figure \ref{BH} but
for models with PHOT configuration and
with inclination of 45$^\circ$ (left-hand panel)
and 20$^\circ$ (right-hand panel).
\label{incline}}
\end{figure*}

NGC 5128 appears very round but it is possible
that it is intrinsically quite flattened.  Dynamical models
such as those of \citet{wil86} and \citet{hui95} conclude
that is indeed flattened, with axis ratios of 1:0.98:0.55
and 1:0.92:0.79. The orbit-based models described above
are all for an edge-on configuration, i.e. assuming
that NGC 5128 is indeed nearly intrinsically round.  We
test the effect of using a model with different inclinations.
We take the case of inclination $i=20^\circ$, which implies
axis ratios of 1:1:0.5, and $i=45^\circ$, which implies
axis ratios of 1:1:0.9.
Figure \ref{incline} presents the results for the PHOT
configuration of the kinematic data, the two-dimensional
distribution of $\chi^2$ as a function of BH mass and $M/L$ ratio.
The best-fit BH mass for $i=20^\circ$ is
$1.5^{+0.3}_{-0.2}\times 10^8$ M$_\odot$
and the best-fit $M/L_K$ ratio is $0.68^{+0.02}_{-0.02}$.
The $M/L_K$ ratio is in good agreement with the edge-on models
but the BH mass is $\sim$30\% smaller.
Previous studies using this technique \citep{geb00,geb03}
have found that
on average, inclination appears to cause a 30\% 
random change in the BH mass, exactly what we find here.
The best-fit BH mass for $i=45^\circ$ is
$1.8^{+0.4}_{-0.4}\times 10^8$ M$_\odot$
and the best-fit $M/L_K$ ratio is $0.53^{+0.04}_{-0.03}$,
a BH mass intermediate between the other two inclinations.
The edge-on and $i=20^\circ$ estimates of the BH mass are 
only different by $2\sigma$ and this
somewhat smaller BH mass does not markedly affect our
conclusions below.  The $\chi^2$ of the edge-on model
is significantly less than either inclined model 
($\Delta\chi^2\sim100$) so we adopt that value for the BH mass.

\vskip 30pt
 
\section{Discussion}

Our BH masses for all modeled inclinations are in good
agreement with the gas dynamical results of \citet{mar01}.
This agreement supports these authors' claim that the 
gas kinematics of NGC 5128 are well described by an 
ordered gas disk and suggests that in such situations,
gas dynamics give reliable estimates for BH masses.
Also, NGC 5128 has the
largest offset from the BH-$\sigma$ correlation ever measured.
The BH of NGC 5128 is five to ten times larger than it should be as
predicted by this correlation. 
\citet{mar03} use the BH mass from \citet{mar01}
to place NGC 5128 (along with $\sim$25 other objects) in the
correlation between BH mass and near-infrared bulge luminosity.
\citet{mar03} find that the spread in this correlation
is similar to that of the BH-$\sigma$ correlation.  NGC 5128
does lie above this relation (i.e. its BH mass is somewhat
high compared to its near-IR bulge luminosity) but it 
is not a striking outlier in this relation.
Apparently, the history of NGC 5128 has caused it to have a
very large BH mass for its velocity dispersion but not an
equally high BH mass compared to its near-IR bulge luminosity.

The remarkably high BH mass for NGC 5128 suggests that its
BH assembled before the host stellar bulge.  There are a few
other observations that suggest this. \citet{wal04} use
molecular gas observations of the host galaxy of a $z=6.42$
quasar to show that it is missing the large stellar bulge
implied by the BH-$\sigma$ and BH-$L_{bulge}$ correlations. 
A large BH is in
place, but there does not appear to be an associated stellar
bulge.  This, along with our result for NGC 5128,
is in sharp difference to the census of BHs
in local galaxies described by \citet{ho04} and others and the
result for QSOs \citep{shi03}.  There are several attractive
theoretical models which explain the observed BH-$\sigma$
correlation \citep{sil98,fab99,kin03} and some also naturally
explain such galaxy properties as color bimodality and
the Faber-Jackson relation \citep{mur04,spr04}.  These models
all use similar feedback arguments to show how an actively
accreting BH can regulate its own growth by expelling gas 
from its immediate vicinity through an outflow or wind.  The wind
becomes stronger as the BH grows, driving away the gas that fuels
the BH growth; the models connect the ultimate size the BH can attain
with the host galaxy characteristics (i.e. $\sigma$).  
These models have been invoked to explain the 
BH-$\sigma$ correlation of normal galaxies; understanding NGC 5128
and its offset from the BH-$\sigma$ relation in the context
of these arguments will provide an important clue for our understanding
of the role of central BHs in galaxy formation.

Why is NGC 5128 such an unusual galaxy in this respect?  The two
most obvious possibilities are the recent major merger it has 
undergone and its AGN activity.  The AGN activity seems to be a
likely culprit.  It is possible that a supermassive black hole
that is currently growing by accreting material could be large
compared to its host galaxy; perhaps the host galaxy has not
had time to catch up to the growth of the BH.  However, there
has been some study of the BH-$\sigma$ correlation in
galaxies with active nuclei and this is not what has been found.  
\citet{geb00b} studied the relationship
between galaxy velocity dispersion and BH mass measured by
reverberation mapping for a sample of active galaxies and
found that these active galaxies are consistent with the
BH-$\sigma$ correlation for quiescent galaxies.  Previously it
had been noted that BH masses measured by reverberation mapping 
in active galaxies fall significantly below the correlation 
between bulge luminosity and BH mass determined from 
spatially resolved kinematics of nearby normal galaxies.
This is interesting in light of our results for NGC 5128.
Our BH mass for NGC 5128 is relatively consistent with the
BH-$L_{bulge}$ relationship but too high for the BH-$\sigma$
correlation, while the reverberation mapping BH masses for
AGN galaxies were too low for the BH-$L_{bulge}$ correlation
but consistent with the BH-$\sigma$ correlation.  Perhaps the
reverberation mapping masses underestimate the true BH mass and
AGN galaxies have larger BH masses for their velocity
dispersions than quiescent galaxies.  We need more BH masses
measured through stellar (or possibly in some cases, gas)
dynamics for active galaxies to further explore this issue.

NGC 5128's merger history presents another possibility which
has been less studied until now.  Perhaps some aspect of
the merging process builds up the central black hole of a
galaxy before later evolution causes the galaxy's velocity
dispersion to catch up.  For instance, if two galaxies with
central supermassive BHs merge, their BHs may merge into
one BH quickly while it may take much longer for the new galaxy
to assume its final dynamical configuration.
If some scenario like this is the case, we have happened to
catch NGC 5128 at this stage in its merger when its 
supermassive BH has this specific relationship to the 
galaxy as a whole.  Little is known about the central
BHs of galaxies that have undergone recent mergers; NGC 5128
is the first such galaxy to have its central BH measured.

It is also possible that both of these aspects are acting in
NGC 5128, of course.  For both of these questions, using
near-infrared kinematics holds promise for making progress.
Many AGN and all recent merger galaxies are significantly
dusty and thus inaccessible to optical spectroscopy.
Using the techniques described here, we can 
reliably measure the kinematics of such galaxies.
NGC 5128 is so enshrouded in dust that even its $K$-band light
shows signs of reddening but this does not hamper our
kinematic measurements or dynamical modeling.
Also, with the untimely death
of STIS on HST, our main method of measuring black holes has been
eliminated and we can no longer rely on our previous techniques.
If we want to continue to explore the connections between black holes
and their host galaxies that HST first uncovered,
the only method currently available is to use stellar kinematics measured
at near-IR wavelengths.  The excellent atmospheric seeing in
the near-IR at good telescope sites, such as with Gemini and GNIRS,
allows us to probe the central
regions of nearby galaxies where the gravitational effects of the black hole
are strongest.  Also, as adaptive optics instrumention becomes
more available, we can push to higher spatial resolution and thus
the sphere of influence of black holes of more distant galaxies.

\acknowledgments{
We thank Bernadette Rodgers for assistance in understanding and
using the GNIRS instrument during the system verification process.
KG and JS gratefully acknowledge the 
support of the Texas Advanced Research
Program and Grant No. 003658-0243-2001.
KG acknowledges NSF CAREER grant AST-0349095.
This publication is based on observations obtained at the Gemini 
Observatory, which is operated by the Association of Universities 
for Research in Astronomy, Inc., under a cooperative agreement
with the NSF on behalf of the Gemini partnership: the National 
Science Foundation (United States), the Particle Physics and 
Astronomy Research Council (United Kingdom), the National Research 
Council (Canada), CONICYT (Chile), the Australian Research Council
(Australia), CNPq (Brazil) and CONICET (Argentina).
This publication makes use of data products from the 
Two Micron All Sky Survey, which is a joint project of the 
University of Massachusetts and the Infrared Processing and 
Analysis Center/California Institute of Technology, funded by the 
National Aeronautics and Space Administration and the National 
Science Foundation.}

\clearpage
\begin{deluxetable}{rcccccccc}
\tablecaption{Gauss-Hermite moments of LOSVDs for NGC 5128\label{momenttable}}
\tablewidth{0pt}
\tablehead{
& \multicolumn{4}{c}{parallel to dust disk, offset} & \multicolumn{4}{c}{perpendicular to dust disk, centered} \\
\colhead{radius} & \colhead{$v$} & \colhead{$\sigma$} & \colhead{h$_3$} & \colhead{h$_4$} & \colhead{$v$} & \colhead{$\sigma$} & \colhead{h$_3$} & \colhead{h$_4$} \\
\colhead{($^{\prime\prime}$)} & (km s$^{-1}$)  & (km s$^{-1}$)  &  &  & (km s$^{-1}$)  & (km s$^{-1}$)  &  &  \\
}
\startdata

0.00  & -12$\pm$10 & 153$\pm$13 &  0.036$\pm$0.053  &  0.056$\pm$0.051  & & & & \\
0.15  &  -4$\pm$9  & 141$\pm$11 & -0.042$\pm$0.036  &  0.054$\pm$0.039  & & & & \\
0.30  & -16$\pm$10 & 156$\pm$9  &  0.020$\pm$0.038  & -0.011$\pm$0.036  & 10$\pm$27 & 233$\pm$44 &  0.013$\pm$0.137  & 0.309$\pm$0.129 \\
0.45  & -21$\pm$9  & 154$\pm$9  &  0.025$\pm$0.035  & -0.007$\pm$0.033  & 14$\pm$19 & 198$\pm$34 & -0.008$\pm$0.074  & 0.217$\pm$0.105 \\
0.60  &  -6$\pm$8  & 142$\pm$9  &  0.001$\pm$0.032  &  0.088$\pm$0.039  & 20$\pm$12 & 174$\pm$14 & -0.024$\pm$0.048  & 0.050$\pm$0.047 \\
0.75  & -12$\pm$8  & 145$\pm$8  &  0.008$\pm$0.029  &  0.025$\pm$0.031  &  8$\pm$6  & 159$\pm$10 &  0.023$\pm$0.035  & 0.020$\pm$0.027 \\
1.05  &   1$\pm$8  & 146$\pm$9  & -0.047$\pm$0.033  &  0.035$\pm$0.033  &  6$\pm$8  & 141$\pm$7  &  0.004$\pm$0.035  & 0.008$\pm$0.024 \\
1.65  &  -7$\pm$7  & 140$\pm$7  & -0.016$\pm$0.025  & -0.004$\pm$0.024  & 10$\pm$5  & 140$\pm$5  & -0.028$\pm$0.022  & 0.009$\pm$0.018 \\
2.70  &  -8$\pm$8  & 139$\pm$8  &  0.009$\pm$0.028  &  0.032$\pm$0.033  &  5$\pm$5  & 142$\pm$6  &  0.030$\pm$0.024  & 0.024$\pm$0.021 \\
3.75  & -11$\pm$7  & 139$\pm$7  &  0.013$\pm$0.027  &  0.006$\pm$0.026  & 11$\pm$5  & 138$\pm$6  & -0.007$\pm$0.027  & 0.029$\pm$0.026 \\
5.70  &  -4$\pm$7  & 136$\pm$7  & -0.013$\pm$0.025  & -0.011$\pm$0.023  & 14$\pm$4  & 131$\pm$5  & -0.002$\pm$0.021  & 0.003$\pm$0.016 \\
8.55  & -20$\pm$7  & 133$\pm$8  & -0.002$\pm$0.029  &  0.020$\pm$0.027  & 14$\pm$5  & 132$\pm$5  &  0.022$\pm$0.022  & 0.016$\pm$0.019 \\
12.60 & -17$\pm$7  & 120$\pm$7  & -0.006$\pm$0.027  & -0.017$\pm$0.020  & 19$\pm$5  & 132$\pm$5  & -0.006$\pm$0.022  & 0.027$\pm$0.022 \\
18.45 & -19$\pm$12 & 130$\pm$12 &  0.022$\pm$0.039  &  0.018$\pm$0.035  & 26$\pm$6  & 139$\pm$6  &  0.005$\pm$0.029  & 0.050$\pm$0.027 \\
27.00 &   9$\pm$19 & 136$\pm$19 &  0.021$\pm$0.059  & -0.012$\pm$0.059  & 19$\pm$6  & 141$\pm$7  & -0.014$\pm$0.028  & 0.022$\pm$0.021 \\
38.85 & -17$\pm$19 & 141$\pm$27 & -0.158$\pm$0.101  &  0.086$\pm$0.116  &  9$\pm$8  & 139$\pm$10 &  0.040$\pm$0.048  & 0.037$\pm$0.044 \\
82.35 &&&&&  19$\pm$22\tablenotemark{a} & 139$\pm$16\tablenotemark{a} & & \\
129.4 &&&&&  29$\pm$23\tablenotemark{a} & 143$\pm$16\tablenotemark{a} & & \\
158.8 &&&&&  34$\pm$19\tablenotemark{a} & 122$\pm$14\tablenotemark{a} & & \\
\enddata
\footnotesize{\tablerefs{(a) \citet{pen04}}}
\end{deluxetable}

\end{document}